\documentstyle{mn}
\newif\ifAMStwofonts
\ifoldfss
  \ifCUPmtlplainloaded \else
    \NewTextAlphabet{textbfit} {cmbxti10} {}
    \NewTextAlphabet{textbfss} {cmssbx10} {}
    \NewMathAlphabet{mathbfit} {cmbxti10} {} % for math mode
    \NewMathAlphabet{mathbfss} {cmssbx10} {} %  "   "    "
  \fi
  \ifAMStwofonts
    \ifCUPmtlplainloaded \else
      \NewSymbolFont{upmath} {eurm10}
      \NewSymbolFont{AMSa} {msam10}
      \NewMathSymbol{\upi}     {0}{upmath}{19}
      \NewMathSymbol{\umu}     {0}{upmath}{16}
      \NewMathSymbol{\upartial}{0}{upmath}{40}
      \NewMathSymbol{\leqslant}{3}{AMSa}{36}
      \NewMathSymbol{\geqslant}{3}{AMSa}{3E}

    \fi
  \fi
\fi % End of OFSS

\ifnfssone
  \newmathalphabet{\mathit}
  \addtoversion{normal}{\mathit}{cmr}{m}{it}
  \addtoversion{bold}{\mathit}{cmr}{bx}{it}
  \newmathalphabet{\mathbfit} % math mode version of \textbfit{..}
  \addtoversion{normal}{\mathbfit}{cmr}{bx}{it}
  \addtoversion{bold}{\mathbfit}{cmr}{bx}{it}
  \newmathalphabet{\mathbfss} % math mode version of \textbfss{..}
  \addtoversion{normal}{\mathbfss}{cmss}{bx}{n}
  \addtoversion{bold}{\mathbfss}{cmss}{bx}{n}
  \ifAMStwofonts
    \ifCUPmtlplainloaded \else
      \UseAMStwoboldmath
      \makeatletter
      \new@mathgroup\upmath@group
      \define@mathgroup\mv@normal\upmath@group{eur}{m}{n}
      \define@mathgroup\mv@bold\upmath@group{eur}{b}{n}
      \edef\UPM{\hexnumber\upmath@group}
      \new@mathgroup\amsa@group
      \define@mathgroup\mv@normal\amsa@group{msa}{m}{n}
      \define@mathgroup\mv@bold\amsa@group{msa}{m}{n}
      \edef\AMSa{\hexnumber\amsa@group}
      \makeatother
      \mathchardef\upi="0\UPM19
      \mathchardef\umu="0\UPM16
      \mathchardef\upartial="0\UPM40
      \mathchardef\leqslant="3\AMSa36
      \mathchardef\geqslant="3\AMSa3E
    \fi
  \fi
\fi % End of NFSS release 1

\ifnfsstwo
  \DeclareMathAlphabet{\mathbfit}{OT1}{cmr}{bx}{it}
  \SetMathAlphabet\mathbfit{bold}{OT1}{cmr}{bx}{it}
  \DeclareMathAlphabet{\mathbfss}{OT1}{cmss}{bx}{n}
  \SetMathAlphabet\mathbfss{bold}{OT1}{cmss}{bx}{n}
  \ifAMStwofonts
    \ifCUPmtlplainloaded \else
      \DeclareSymbolFont{UPM}{U}{eur}{m}{n}
      \SetSymbolFont{UPM}{bold}{U}{eur}{b}{n}
      \DeclareSymbolFont{AMSa}{U}{msa}{m}{n}
      \DeclareMathSymbol{\upi}{0}{UPM}{"19}
      \DeclareMathSymbol{\umu}{0}{UPM}{"16}
      \DeclareMathSymbol{\upartial}{0}{UPM}{"40}
      \DeclareMathSymbol{\leqslant}{3}{AMSa}{"36}
      \DeclareMathSymbol{\geqslant}{3}{AMSa}{"3E}
    \fi
  \fi
\fi % End of NFSS release 2

\ifCUPmtlplainloaded \else
  \ifAMStwofonts \else % If no AMS fonts
    \def\upi{\pi}
    \def\umu{\mu}
    \def\upartial{\partial}
  \fi
\fi

\title[The H{\normalsize \it I} content of Fornax dwarf elliptical
galaxies]{The H{\sc i} content of Fornax dwarf elliptical
galaxies:~FCC032 and FCC336} \author[P. Buyle, S. De Rijcke,
D. Michielsen, M. Baes, H. Dejonghe] {P. Buyle$^1$
\thanks{corresponding author: Pieter.Buyle@UGent.be}, S. De Rijcke$^1$
\thanks{Postdoctoral Fellow of the Fund for Scientific Research -
Flanders (Belgium)(F.W.O)}, D. Michielsen$^1$, M. Baes$^{1,2}$,
H. Dejonghe$^1$ \\ $^1$ Sterrenkundig Observatorium, Universiteit
Gent, Krijgslaan 281, S9, B-9000, Gent, Belgium\\ $^2$ European
Southern Observatory, Casilla 19001, Santiago 19, Chile}
\date{Accepted 1988 December 15.  Received 1988 December 14; in
original form 1988 October 11}

\pagerange{\pageref{firstpage}--\pageref{lastpage}}
\pubyear{1994}

\begin{document}

\maketitle

\label{firstpage}

\begin{abstract}
We present H{\sc i} 21~cm line observations, obtained with the
Australia Telescope Compact Array, of two dwarf elliptical galaxies
(dEs) in the Fornax cluster:~FCC032 and FCC336. The optical positions
and velocities of these galaxies place them well within the Fornax
cluster. FCC032 was detected at the 3$\sigma$ significance level with
a total H{\sc i} flux density of 0.66$\pm$0.22~Jy~km~s$^{-1}$ or an
H{\sc i} mass of 5.0$\pm$1.7$\times 10^7 h_{75}^{-2}\,M_\odot$. Based
on our deep H$\alpha+[{\rm N{\sc ii}}]$ narrow-band images, obtained
with FORS2 mounted on the VLT, this dE was already known to contain
600-1800 $h_{75}^{-2}M_\odot$ of ionised Hydrogen (depending on the
relative strengths of the H$\alpha$ and [N{\sc ii}] emission
lines). Hence, this is the first study of the complex, multi-phase
interstellar medium of a dE outside the Local Group. FCC336 was
detected at the same significance level: 0.37$\pm$0.10~Jy~km~s$^{-1}$
or a total H{\sc i} mass of 2.8$\pm$0.7$\times 10^7
h_{75}^{-2}\,M_\odot$. Using a compilation of H{\sc i} data of dwarf
galaxies, we find that the observed high H{\sc i}-mass boundary of the
distribution of dIrrs, BCDs, and dEs in a $\log ( L_B )$ versus $\log
(M_{\rm H{\sc i}})$ diagram is in good agreement with a simple
chemical evolution model with continuous star formation. The existence
of many gas-poor dEs (undetected at 21~cm) suggest that the
environment (or more particularly, a galaxy's orbit within a cluster)
also plays a crucial role in determining the amount of gas in
present-day dEs. E.g., FCC032 and FCC336 are located in the sparsely
populated outskirts of the Fornax cluster. This is in agreement with
H{\sc i} surveys of dEs in the Virgo Cluster and an H$\alpha$ survey
of the Fornax Cluster, which also tend to place gas-rich dwarf
galaxies in the cluster periphery.
\end{abstract}


\section{Introduction}

One would not expect that dwarf elliptical galaxies (dEs) in dense
environments contain a significant interstellar medium (ISM). Several
arguments support this statement. Supernova-explosions are able to
transfer enough energy to the ISM to heat it above the escape velocity
in the least massive dwarfs \cite{mo97}. Alternatively, the frequent
high-speed interactions with giant cluster-members to which a small
late-type disk galaxy is subjected can transform it into a gasless
spheroidal dE-like object. This ``galaxy harassment'' process
\cite{mo96} induces a dramatic morphological evolution on a time-span
of about 3~Gyr. Moreover, hydrodynamical simulations of dwarf galaxies
moving through the hot, thin intergalactic medium in clusters
\cite{mb00,mi04b} or groups \cite{ma03} show that ram-pressure
stripping can completely remove the ISM of a dwarf galaxy less massive
than $10^9 M_\odot$ within a few 100 Myrs. A quite different point of
view on the origin of dEs comes to the same conclusion. If dEs are
related to other dwarf galaxies such as Blue Compact Dwarfs (BCDs) or
dwarf irregular galaxies (dIrrs), the ``fading model'' conjectures
that star-forming dwarf galaxies will fade and reach an end-state
similar to present-day dEs after they have used up their gas supply
and star-formation has ended \cite{ma99}. Interactions may have sped
up the gas-depletion process \cite{vz04}, explaining both the
abundance of dEs and the paucity of BCDs/dIrrs in high-density
environments.

For all these reasons, dEs in dense environments were generally
thought to be virtually gas-depleted systems. However, evidence is
building up that at least some dEs have retained part of their gas. In
their multi-wavelength study of the Local Group dwarf galaxies, Young
\& Lo presented VLA H{\sc i} observations of NGC147, NGC185, and
NGC205 \cite{yl96,yl97}. These were the first observations that
painted a detailed picture of the complex, multi-phase interstellar
medium (ISM) of the most nearby representatives of the class of the
dEs \cite{fb94}. While NGC147 was not detected with a 3~$\sigma$ mass
upper limit of $3 \times 10^3\,M_\odot$ for an 8~km~s$^{-1}$ velocity
width, NGC205 was found to contain $4.3 \times 10^5\,M_\odot$ of
neutral hydrogen and the total H{\sc i} mass of NGC185 was estimated
at $1.0 \times 10^5\,M_\odot$. The neutral ISM of both detected
galaxies turned out to be very clumpy, making a meaningful
determination of their velocity fields rather difficult. Still, the
stars and H{\sc i} gas in NGC205 seem to have different rotation
velocities while in NGC185, neither the H{\sc i} or the stars show
significant rotation \cite{sp02}. Single-dish observations of
$^{12}$CO emission provide evidence that the molecular and atomic gas
are kinematically linked. NGC205 was not detected on H$\alpha+$[NII]
narrow-band images while NGC185 contains an extended emission region,
about 50~pc across \cite{yl96,yl97}.

More recently, H{\sc i} surveys of the Virgo Cluster dE population
(see Conselice~et al.~(2003) and references therein) have shown that
roughly 15\% of the dEs contain a neutral ISM. The detected H{\sc i}
masses range between 0.03 and $1.22 \times 10^9\,M_\odot$. Processes
that remove gas, such as galaxy interactions and ram-pressure
stripping \cite{vo03,ma03,mi04b,rh05}, act most vigorously near the
cluster center. Accordingly, the gas-rich dwarf galaxies in the Virgo
cluster tend to have positions towards the outskirts of the cluster,
suggesting that they are recent acquisitions of the cluster or are
moving on orbits that avoid the cluster center. In a spectroscopic
survey of the Fornax Cluster, Drinkwater et al.~(2001) discovered
H$\alpha$ emission in about 25\% of the dEs. Again, most of these
galaxies lie towards the outskirts of the cluster, while dEs near the
center of the cluster are generally devoid of ionized gas.

\begin{figure}
\vspace*{7.25cm}
\special{hscale=85 vscale=85 hsize=250 vsize=250
hoffset=-20 voffset=-310 angle=0 psfile="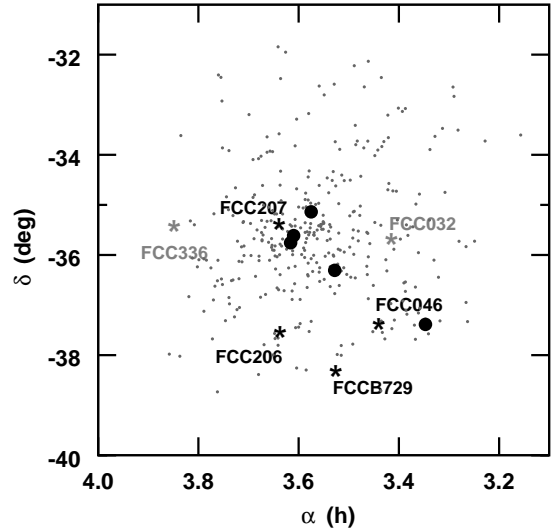"}
\caption{Positions of gas-rich dEs within the Fornax Cluster. Small
dots indicate the positions of 340 galaxies in the Fornax Cluster
Catalog. Large dots indicate the positions of galaxies brighter than
$M_B < -20$~mag. Black asterisks indicate the positions of FCC046,
FCC206, FCC207, and FCCB729. These dEs contain an ionised ISM (De
Rijcke et al.~(2003), Michielsen et al.~(2004)). The positions of the
H{\sc i} detected Fornax dEs presented in this paper, FCC032 and
FCC336, are indicated with a grey asterisk. FCC032 was detected both
in H$\alpha$ and H{\sc i} 21~cm emission.\label{cat}}
\end{figure}

In this paper, we present new H{\sc i} 21~cm line observations of two
dEs, obtained with the Australia Telescope Compact Array (ATCA). With
optical systemic velocities $v_{\rm sys} = 1318 \pm 26$~km/s (FCC032)
and $v_{\rm sys} = 1956 \pm 67$~km/s (FCC336) \cite{drink01}, these
dEs are bona fide members of the Fornax Cluster, located in the
sparsely populated outskirts of the cluster (see Fig. \ref{cat}). In
section \ref{hi}, we present our H{\sc i} observations, followed by a
discussion of our results in section \ref{disc}. We summarize our
conclusions in section \ref{conc}.

\section{H{\sc i} observations} \label{hi}

We have used the Australia Telescope Compact Array on 20 and 23
December 2004 to observe two dEs in the Fornax cluster. We preferred
interferometry observations above single-dish observations to avoid
confusion with other galaxies that can be located within the large
beam, which is a common nuisance in crowded environments such as the
Fornax Cluster. The observations were made during night time to avoid
solar RFI. We used the ATCA in the 1.5D configuration, with baselines
ranging from 107~m to 4439~m. To be able to detect H{\sc i} emission
in three independent channels and since both sources had an estimated
velocity width of about 50~km\ s$^{-1}$, we selected a correlator
setup that yielded 512 channels of width 15.6~kHz. To increase the
signal-to-noise ratio the data were on-line Hanning smoothed which
resulted in a velocity resolution of $6.6$~km\ s$^{-1}$. At the start
of each observation we observed the source 1934-638 as primary
calibrator for 15 minutes. The source 0332-403 was observed every 40
minutes for 5 minutes as a secondary calibrator. The total integration
time (including calibration) for each galaxy was 12h. The usual data
reduction steps (phase, amplitude and bandpass calibration) were
performed with the MIRIAD package \cite{sa95}, the standard ATCA data
analysis program. We subtracted the continuum by performing a first
order fit to the visibilities over the line-free channels which were
not affected by the edge effects of the band (selected in advance by
eye). The data cubes were created by using natural weighting and were
subsequently smoothed with a Gaussian beam of 1$' \times1'$ (which
corresponds to the optical spatial radius of our sources) and off-line
Hanning smoothed to increase the signal-to-noise. The final data cubes
had a spectral resolution of $13.3$ km\ s$^{-1}$. Due to the faintness
of these objects, we did not attempt a deconvolution of our images.

\subsection{FCC032}

FCC032 was selected as a suitable target because it was known to
contain a sizable ionised ISM \cite{mi04}. This galaxy harbors a large
ionised gas complex, about 850~pc across, containing star-formation
regions and bubble-shaped gaseous filaments, probably supernova
remnants. This offers strong evidence for recent or ongoing star
formation in this dE. 

\begin{figure}
\vspace*{7.25cm} \special{hscale=63 vscale=58 hsize=550 vsize=500
hoffset=-30 voffset=-120 angle=0 psfile="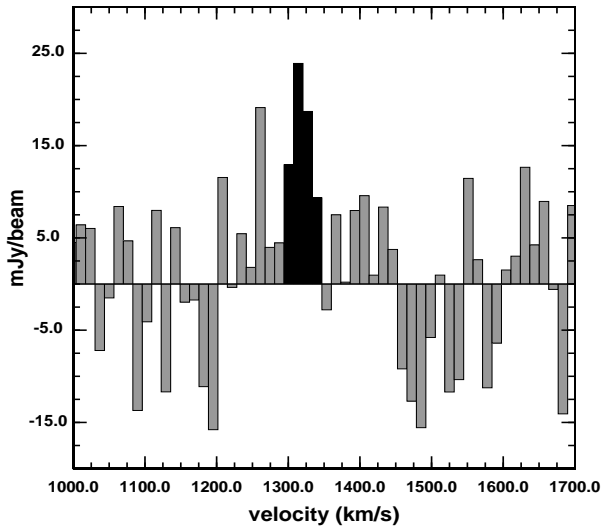"}
\caption{H{\sc i} profile of FCC032, obtained by summing the flux
within a $2' \times 2'$ box around the central radio position of the
galaxy. The channels containing the 21~cm emission of FCC032 are
colored in black. \label{spec032}}
\end{figure}

\begin{figure}
\vspace*{7.25cm} \special{hscale=63 vscale=58 hsize=550 vsize=500
hoffset=-10 voffset=-30 angle=0 psfile="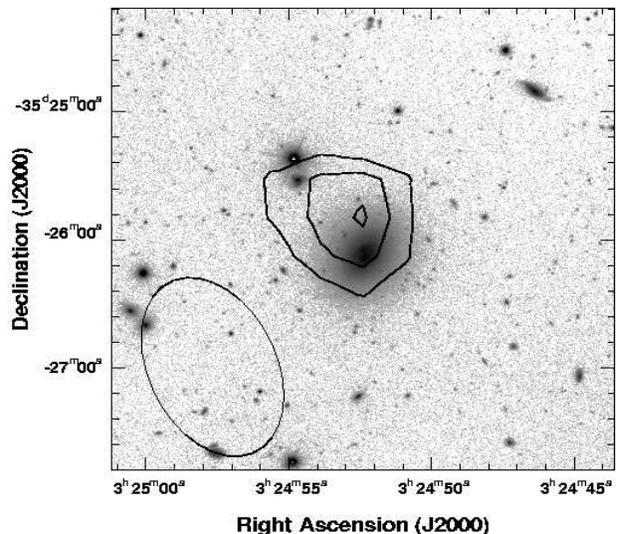"}
\caption{Total H{\sc i} intensity map (contours) of FCC032 on top of
an R-band grey-scale image (obtained with FORS2, VLT). The contours
correspond to the 2$\sigma_N^h$, 2.5$\sigma_N^h$, and the
3$\sigma_N^h$ levels with $\sigma_N^h=0.11$~Jy/beam~km~s$^{-1}$ (see
eq. \ref{sumsigma}). The beam is plotted in the lower-left
corner. \label{total032}}
\end{figure}

Our final data cube had a synthesised beam of $87.56'' \times 55.55''$
due to the smoothing, resulting in a noise of 4~mJy/beam. These data
cubes were inspected by eye for emission by plotting them over an
optical image. The data were clipped at 1.5$\sigma$=6 mJy/beam. Weak
emission near the optical center of the galaxy was found in 4 adjacent
channels located around the optical systemic velocity. To derive a
final spectrum (containing all detected HI), we summed the flux within
a $2' \times 2'$ box centered on the central radio position of FCC032
which was derived from the total HI intensity map (see
Fig.\ref{total032}). This resulted in a detection of the galaxy near
the optical systemic velocity of 1318 km\ s$^{-1}$, spread over 4
channels (see Fig. \ref{spec032}).

We fitted a Gaussian to the global H{\sc i} profile and adopted the
method of Verheijen \& Sancisi~(2001) to measure the systemic velocity
and line widths at the 20\% and 50\% levels (corrected for broadening
and random motions). We applied the instrumental correction given by
the formulae
\begin{equation}
W_{20}=W_{20\textrm{,obs}}-35.8\left[\sqrt{1+(R/23.5)^2}-1\right],
\end{equation}
\begin{equation}
W_{50}=W_{50\textrm{,obs}}-23.5\left[\sqrt{1+(R/23.5)^2}-1\right],
\end{equation}
with $R$ the velocity resolution or $13.3$ km\ s$^{-1}$ in our
case. This correction is based on approximating the edges of the
global profile, which are mostly due to turbulent motion, with a
Gaussian of dispersion $10$ km\ s$^{-1}$. The results show a maximum
flux of 24~mJy/beam at a velocity of 1318~km~s$^{-1}$, which we
adopt as the radio systemic velocity of FCC032 and which is in
excellent agreement with the optically derived systemic velocity
\cite{drink01}. For the widths at 20\% and 50\% we found respectively
52~km~s$^{-1}$ and 34~km~s$^{-1}$.

\begin{figure}
\vspace*{7.25cm} \special{hscale=63 vscale=58 hsize=550 vsize=500
hoffset=-10 voffset=-20 angle=0 psfile="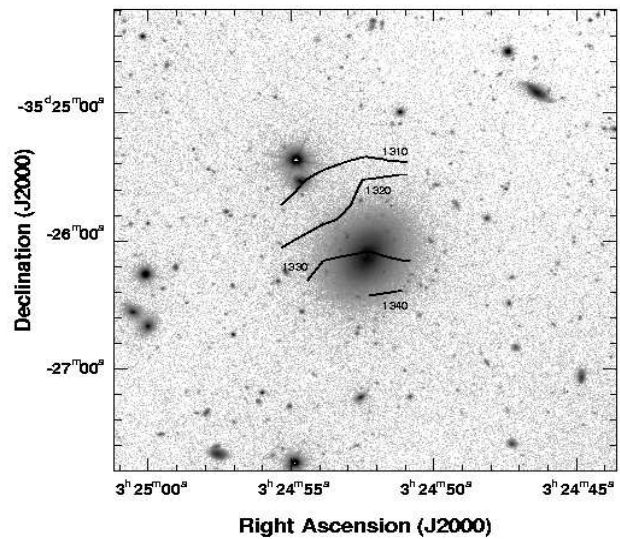"}
\caption{H{\sc i} intensity-weighted velocity field of FCC032
(contours) on top of an R-band grey-scale image (obtained with FORS2,
VLT). The contours are labeled with the velocity in
km~s$^{-1}$. \label{map032}}
\end{figure}

By summing the flux within the 4 channels that contain the 21~cm
emission of FCC032, we created a total H{\sc i} intensity map (see
Fig. \ref{total032}). The noise at a certain position in the total
intensity map can be calculated by means of the formula \cite{vs01}:
\begin{equation}
\sigma_N^h=\left(\frac{N}{2}-\frac{1}{8}\right)^{\frac{1}{2}}
\frac{4}{\sqrt{6}}\sigma^h
\label{sumsigma}
\end{equation}
where $N$ stands for the number of channels that have been added at
that position and with $\sigma^h$ the noise in the Hanning smoothed
channel maps (4~mJy/beam or 53.2~mJy/beam~km~s$^{-1}$). Not all
channels contribute to the flux at a given position in the total
intensity map so $N$ is not the same everywhere. Typically, $N=3$ in
this case. This yields an average spatial rms error $\sigma_N^h
\approx 0.11$~Jy/beam~km~s$^{-1}$. We rebinned the channels plotted in
Fig. \ref{spec032} by taking together 4 adjacent channels, such that
all the 21~cm flux of FCC032 ends up in one single bin,
53.2~km~s$^{-1}$ wide, and then calculated the rms noise from the
other bins. This way, we find a total velocity integrated H{\sc i}
flux density of 0.66$\pm$0.22~Jy~km~s$^{-1}$ and calculated a total
estimated H{\sc i} mass of 5.0$\pm$1.7$\times 10^7
h_{75}^{-2}\,M_\odot$, based on the $W_{20}$ velocity width by means
of the formula:
\begin{equation}
M_{\rm {H{\sc i}}}=2.36\times10^5\ D^2 \ \int S(v)dv\,\, M_\odot
\label{massa}
\end{equation}
with $D=18~h_{75}^{-2}~$Mpc the distance to the Fornax cluster and
$\int S(v)dv$ the total flux density in units of Jy~km~s$^{-1}$.  The
H{\sc i} intensity-weighted velocity field of FCC032 is presented in
Fig. \ref{map032}. A velocity gradient can be observed, suggesting
rotation.

\subsection{FCC336}

\begin{figure}
\vspace*{7.25cm} \special{hscale=63 vscale=58 hsize=550 vsize=500
hoffset=-30 voffset=-120 angle=0 psfile="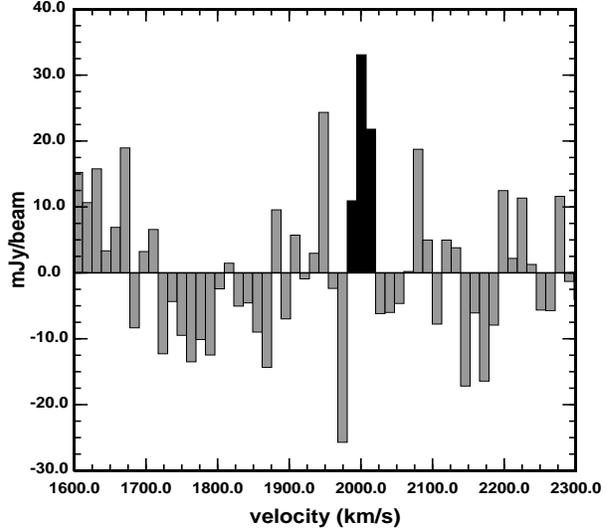"}
\caption{H{\sc i} profile of FCC336, obtained by summing the flux
within a $2' \times 2'$ box around the central radio position of the
galaxy. The channels containing the 21~cm emission of FCC336 are
colored in black. \label{spec336}}
\end{figure}

\begin{figure}
\vspace*{7.25cm} \special{hscale=63 vscale=58 hsize=550 vsize=500
hoffset=-20 voffset=-120 angle=0 psfile="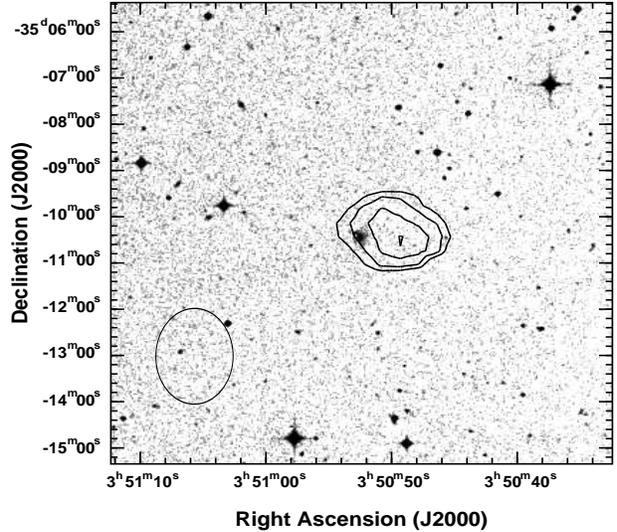"}
\caption{Total H{\sc i} intensity map (contours) of FCC336 on top of
an optical SDSS image (grey-scale). The contours correspond to the
2$\sigma_N^h$, 2.5$\sigma_N^h$, 3$\sigma_N^h$, and 3.3$\sigma_N^h$
levels, with $\sigma_N^h = 0.11$~Jy/beam~km~s$^{-1}$ (see
eq. \ref{sumsigma}). The beam is plotted in the bottom-left
corner. \label{total336}}
\end{figure}

FCC336 was selected according to its position in the outskirts of the
Fornax cluster and thus, in accordance with the ram-pressure stripping
theory, could be expected to harbor a relatively large amount of
neutral hydrogen. Smoothing was again applied to the data cube to
increase the signal-to-noise ratio, resulting in a synthesised beam of
$110.60'' \times 82.13''$ and a noise of 4~mJy/beam. In a similar way
as for FCC032, we found weak emission in three channels near the
optical systemic velocity. We summed the flux within a $2' \times 2'$
box centered on the central radio position of the galaxy, which again
was derived by the total HI intensity map, (see Fig. \ref{total336})
in order to create a global H{\sc i} profile of the galaxy. We found a
strong intensity peak of 34.4~mJy/beam at a velocity of
2004~km~s$^{-1}$ (see Fig. \ref{spec336}), which is in fair agreement
with the optical systemic velocity of 1956 $\pm$ 67~km~s$^{-1}$
\cite{drink01}. A total H{\sc i} intensity map shows that the
emission, allowing for the coarse resolution of the H{\sc i} map,
coincides spatially with the optical image of the galaxy (see
Fig. \ref{total336}). We fit a Gaussian to the global H{\sc i} profile
and find a velocity width of 33~km~s$^{-1}$ and 22~km~s$^{-1}$ at
respectively the 20\% and 50\% levels. We measure a total H{\sc i}
flux density of 0.37$\pm$0.10~Jy~km~s$^{-1}$ , which, by means of
formula (\ref{massa}), corresponds to an estimated total H{\sc i} mass
of 2.8$\pm$0.7 $\times 10^7 h_{75}^{-2}\,M_\odot$. Here, we use the
summed rms within the $2' \times 2'$ box and the $W_{20}$ velocity
width. Due to our velocity resolution of 13.3~km~s$^{-1}$ and the very
small velocity width of FCC336, we did not attempt constructing an
H{\sc i} intensity-weighted velocity field.

\section{Discussion} \label{disc}

\begin{figure}
\vspace*{7.5cm} \special{hscale=65 vscale=65 hsize=550 vsize=500
hoffset=-32 voffset=-135 angle=0 psfile="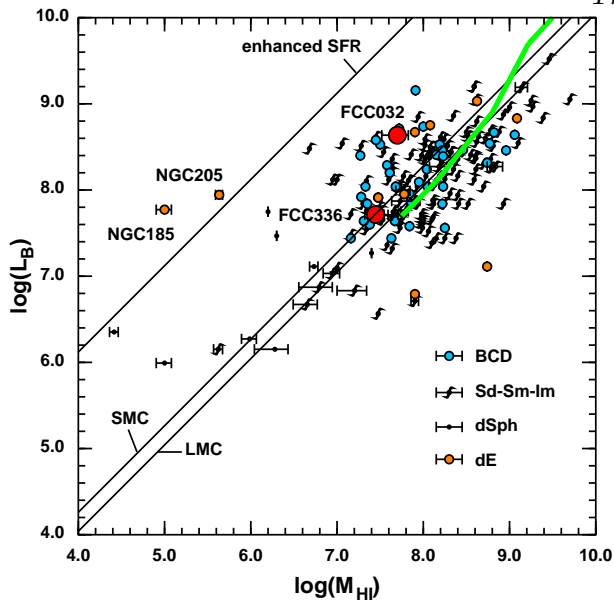"}
\caption{B-band luminosity (in $L_{\odot,B}$) versus H{\sc i} mass (in
$M_\odot$). Large white circles indicate FCC032 and FCC336; small
white circles correspond to Virgo dwarf ellipticals, except for the
two leftmost white circles, which are NGC185 and NGC205. The small
grey dots are Virgo Cluster BCDs. Late-type dwarfs (Sd-Sm-Im) in the
Local Group, the Virgo Cluster and in the field are indicated by
spiral symbols. Black dots correspond to Local group and field dwarf
spheroidals. The full lines trace the sequences predicted by simple
chemical evolution models for various star-formation efficiencies; the
grey curve traces the $\log(L_B)$ vs. $\log(M_{\rm H{\sc i}})$
relation of late-type galaxies predicted by semi-analytical models of
galaxy formation via hierarchical merging in a $\Lambda$CDM universe
(see section \ref{disc}). We stress the fact that only {\em
detected galaxies have been included in this figure. Many dEs,
however, are too gas-poor to have been detected in H{\sc i} at the
distances of the Virgo and Fornax clusters. H{\sc i} studies in these
clusters typically have mass-limits of the order of $10^7
M_\odot$. Only inside the Local Group have dEs and dSphs with a gas
content as low as $\sim 10^5 M_\odot$ been detected. It should
therefore be kept in mind that many dEs reside in the upper left part
of this figure but have not yet been detected.}
\label{plotje}}
\end{figure}

In Fig. \ref{plotje}, we plot the B-band luminosities of FCC032 and
FCC336 versus their H{\sc i} masses, along with the Virgo dEs compiled
by Conselice~et al.~(2003), the Local Group dEs NGC185 and NGC205
\cite{yl96,yl97}, the Local Group dwarf spheroidals (dSphs) and dwarf
irregulars (dIrrs), taken from Mateo~(1998), field dwarf irregulars
and spheroidal galaxies, taken from Roberts et al.  (2004), and Virgo
Cluster Blue Compact Dwarfs (BCDs) and late-type dwarf galaxies
(Sd-Sm-Im), taken from Gavazzi~et al.~(2005) and Sabatini~et
al.~(2005). We included only those galaxies in this diagram that were
actually detected at 21~cm. Many dEs, however, are too gas-poor to
have been detected in H{\sc i} at the distances of the Virgo and
Fornax clusters. H{\sc i} studies in these clusters typically have
mass-limits of the order of $10^7 M_\odot$. Only inside the Local
Group have dEs and dSphs with a gas content as low as $\sim 10^5
M_\odot$ been detected. It should therefore be kept in mind that many
dEs reside in the upper left part of this figure that have so far
evaded detection.

Clearly, BCDs and dIrrs seem to trace a sequence, defined roughly by
the relation $\log(L_B) \approx \log(M_{\rm H{\sc i}})$. All these
galaxies reside either in the Virgo cluster or in the Local group,
with secure distance estimates \cite{ma98}. This makes us confident
that this sequence is not a spurious result of the distance-dependence
of both $\log(L_B)$ and $\log(M_{\rm H{\sc i}})$.  The Local Group
dSphs and the Local Group dEs NGC185 and NGC205 (both are satellites
of M31) deviate from this sequence by being gas
deficient. Non-detections at 21~cm were not plotted in this
figure. Hence, many undetected gas-poor dwarf galaxies (like NGC147,
with a 3$\sigma$ upper limit of $3 \times 10^3\,M_\odot$ for $M_{\rm
H{\sc i}}$ (Young \& Lo, 1997), or the 20 Virgo dEs that were not
detected by Conselice~et al.~(2003), with mass upper limits of $\sim 8
\times 10^6\,M_\odot$) are expected to occupy the left part of the
diagram. Therefore, the $L_B$ vs. $M_{\rm H{\sc i}}$ sequence of
gas-rich dwarf galaxies in Fig. \ref{plotje} is best seen as a
boundary, enclosing the most H{\sc i}-rich galaxies while many
gas-poor dwarfs (like dEs and dSphs) lie significantly to the left of
this sequence.

\subsection{Chemical evolution of dwarf galaxies}

In order to interpret this diagram, we overplotted the observed data
points with theoretical predictions for the $\log(L_B)$
vs. $\log(M_{\rm H{\sc i}})$ relation, based on the analytical models
of Pagel~\&~Tautvai{\v s}ien\.e (1998) for the chemical evolution of
the Large and Small Magellanic Clouds. In the formalism of these
models, galaxies are formed by the infall of pristine gas. Stars are
born at a rate proportional to the gas mass and supernova explosions
eject gas at a rate proportional to the star-formation rate (SFR). The
build-up of the elemental abundances is calculated using the
delayed-recycling approximation in order to include the contribution
of SNIa. We used the B-band mass-to-light ratios of single-age $t$,
single-metallicity $Z$ stellar populations (or SSPs), denoted by
$\Upsilon^*_B(Z,t)$, presented by Vazdekis~et al.~(1996), in
order to calculate the present-day $L_B/M_{\rm H{\sc i}}$ ratio as
\begin{equation}
\frac{L_B}{M_{\rm H{\sc i}}}(T) = \frac{1}{g(T)} \int_0^T \frac{\omega
g(T-t) }{ \Upsilon^*_B(Z(T-t),t)} dt
\end{equation}
with $T= 13~Gyr$ the assumed age of the galaxies, $g(t)$ the gas mass
at time $t$, and $\omega$ the star-formation efficiency (or the
inverse time-scale for star formation). The time-dependence of $g$ and
$Z$ is taken from Pagel~\&~Tautvai{\v s}ien\.e (1998) (their equations
(6), (8), (11), and (14)). The rightmost curve in Fig. \ref{plotje},
labeled with ``LMC'', corresponds to parameter values fine-tuned to
reproduce the elemental abundances observed in the LMC (star-formation
efficiency $\omega=0.18$~Gyr$^{-1}$ and outflow parameter $\eta = 1$);
the middle curve, labeled with ``SMC'', corresponds to the SMC
(star-formation efficiency $\omega=0.115$~Gyr$^{-1}$ and outflow
parameter $\eta = 2$). Cleary, these simple models nicely reproduce
the observed locus of the gas-rich dwarf late types, BCDs, and
dEs. The green curve in Fig. \ref{plotje} traces the $\log(L_B)$
vs. $\log(M_{\rm H{\sc i}})$ relation of late-type galaxies predicted
by semi-analytical models (SAMs) of galaxy formation via hierarchical
merging in a $\Lambda$CDM universe \cite{yn04}. SAMs make use of
a Monte-Carlo technique to construct the hierarchical merger tree that
leads up to the formation of a galaxy of a given mass. They moreover
contain prescriptions for star-formation, energy feedback from
supernova explosions, gas cooling, tidal stripping, dust extinction,
and the dynamical response to starburst-induced gas ejection. Despite
the inevitable oversimplifications in the description of immensely
complex processes such as star formation, they are able to account
pretty well for many observed properties of galaxies. The green curve
in Fig. \ref{plotje} indicates the amount of cold gas present in
simulated galaxies that were classified as late-types (see also Fig. 4
in Nagashima \& Yoshii, 2004).

Thus, it seems that the locus of gas-rich Sd-Sm-Im galaxies and
BCDs in Fig. \ref{plotje} can be reproduced quite satisfactorily by
chemical evolution models of isolated galaxies in which slow star
formation does not exhaust all available gas within a Hubble time.

\subsubsection{Enhanced star-formation rate}

Using the same formalism, one can produce more gas-poor systems by
raising the star-formation efficiency $\omega$. For instance,
interactions may have sped up the gas-depletion process
\cite{vz04,sa05}, explaining both the abundance of gas-poor dEs and
the paucity of gas-rich BCDs/dIrrs in high-density environments.
E.g., the leftmost curve in Fig. \ref{plotje}, labeled with ``enhanced
SFR'', corresponds to a model with a star-formation efficiency that is
a factor of 3 higher than in the LMC model (star-formation efficiency
$\omega=0.54$~Gyr$^{-1}$ and outflow parameter $\eta = 1$). This way,
one can reproduce the locus of gas-poor dwarf galaxies such as the
Local Group dwarf dSphs. All models presented in Fig. \ref{plotje}
have luminosity-weighted mean metallicities in the range
[Fe/H$]=-0.65$ to $-0.4$. The models with a high SFR consume their gas
reservoir in a strong starburst at an early epoch, when the gas was
not yet enriched with metals, and hence, even though they form stars
more efficiently, do not have significantly higher mean metallicities
than the low SFR models (although they do contain a sprinkling of
recently formed metalrich stars which are absent in low SFR
models). As dicussed in Grebel~et al.~(2003), dSphs indeed have more
metalrich red giants than dIrrs and show evidence for a more vigorous
early enrichment than dIrrs.

\subsubsection{Enhanced gas-ejection efficiency}

Raising the gas-ejection efficiency can also enhance the $L_B/M_{\rm
H{\sc i}}$ ratio although, at the same time, it significantly lowers
the mean metallicity by effectively terminating further star-formation
after the first star-forming event (e.g. making the gas-ejection by
supernovae 30 times as efficient as in the Magellanic Clouds leads to
$L_B/M_{\rm H{\sc i}} \approx 3.3$ and [Fe/H$] \approx
-1.6$). However, the star-formation histories of the Local Group dSphs
seem rather continuous (with the exception of the Carina dSph) and
show no evidence for major starbursts which would be able to expell
significant amounts of gas \cite{gr00}. On the other hand, as argued
by Ferrara~\&~Tolstoy (2000), MacLow \& Ferrara (1999), and De Young
\& Heckman (1994), a centralised star-burst event in a round galaxy is
much more efficient at transfering energy to the ISM (and hence at
expelling gas out of a galaxy) than a similar star-burst in a disk
galaxy since, in the latter case, the hot supernova-driven gas can
break through the disky ISM along the minor axis. Hence, this suggests
that rotationally flattened galaxies, such as dwarf late-types, indeed
have lower gas-ejection efficiencies than dEs and dSphs. The
metallicity-flattening relation observed in dEs \cite{bb02}, with more
flattened dEs tending to be less metal-rich, seems to indicate that in
flattened dEs, individual supernova-explosions or star-formation sites
are better able to eject the hot, enriched gas via a chimney
perpendicular to the disk, without appreciably affecting the
surrounding ISM, than in round dEs. This leads to flattening as a
second parameter in controlling a dE's metallicity besides total mass.

\subsubsection{External gas-removal mechanisms}

Hydrodynamical simulations of dwarf galaxies moving through the hot,
rarefied intracluster medium show that ram-pressure stripping can
completely remove the ISM of a low-mass dwarf galaxy
\cite{mb00,vo03,ma03,mi04b}. Interactions and ram-pressure stripping
are most efficient at removing gas from galaxies near the cluster
center. Indeed, the gas-rich dwarf galaxies in the Virgo cluster tend
to have positions towards the outskirts of the cluster
(e.g. Conselice~et al.~(2003)), suggesting that they are recent
acquisitions of the cluster or are moving on orbits that avoid the
cluster center. In a spectroscopic survey of the Fornax Cluster,
Drinkwater et al.~(2001) discovered H$\alpha$ emission in about
25\% of the dEs. Again, most of these galaxies lie towards the cluster
periphery, while dEs near the center of the cluster are generally
devoid of ionized gas. Likewise, both H{\sc i}-rich dEs presented in
this paper are located in the sparsely populated outskirts of the
Fornax cluster (see Fig. \ref{cat}).

\section{Conclusions} \label{conc}

Based on our H{\sc i} 21~cm observations of dEs in the Fornax Cluster
and on H{\sc i} 21~cm observations of dEs, BCDs, and late-type dwarf
galaxies in the Virgo Cluster \cite{co03,ga05}, and the Local Group
dwarfs \shortcite{ma98}, we conclude that the gas-content of the most
gas-rich dwarf galaxies is consistent with a continuous, slow
star-formation history. After one Hubble time, these galaxies still
have a large gas reservoir left and roughly trace a sequence defined
by the relation $\log(L_B) \approx \log( M_{\rm H{\sc i}} )$. However,
the majority of the dwarf spheroidals and dwarf ellipticals contain
significantly less gas than predicted by this relation. External
gas-removal mechanisms such as a star-formation rate enhanced by
gravitational interactions or ram-pressure stripping can account very
well for the existence of these gas-poor systems.  Such external
mechanisms will act most vigorously in high-density environments,
offering a natural explanation for the trend for H{\sc i} mass to
increase with distance from the nearest massive galaxy \cite{gr00},
and the fact that gas-rich dEs are observed predominantly in the
outskirts of clusters.

\section*{Acknowledgments}

We wish to thank C. De Breuck for his kind help and advice and the
anonymous referee for the very helpful suggestions. DM and BP wish to
thank the Bijzonder OnderzoeksFonds (BOF) of Ghent University for
financial support. This paper is based on data obtained with the
Australia Telescope Compact Array (ATCA). The Australia Telescope is
funded by the Commonwealth of Australia for operation as a National
Facility managed by CSIRO. This research has made use of the NASA/IPAC
Extragalactic Database (NED) which is operated by the Jet Propulsion
Laboratory, California Institute of Technology, under contract with
the National Aeronautics and Space Administration.

\bsp \label{lastpage} \end{document}